# How close can one approach the Dirac point in graphene experimentally?

*Alexander S. Mayorov,*[*,†] *Daniel C. Elias,*[†] *Ivan S. Mukhin,*[‡] *Sergey V. Morozov,*[§] *Leonid A. Ponomarenko,*[†] *Kostya S. Novoselov,*[†] *A. K. Geim,*[†,‡] *Roman V. Gorbachev*[†,‡]

[†]School of Physics & Astronomy, University of Manchester, Manchester M13 9PL, UK

[‡]Manchester Centre for Mesoscience & Nanotechnology, Manchester M13 9PL, UK

[§]Institute for Microelectronics Technology, 142432 Chernogolovka, Russia

The above question is frequently asked by theorists who are interested in graphene as a model system, especially in context of relativistic quantum physics. We offer an experimental answer by describing electron transport in suspended devices with carrier mobilities of several $10^6$ cm$^2$V$^{-1}$s$^{-1}$ and with the onset of Landau quantization occurring in fields below 5 mT. The observed charge inhomogeneity is as low as $\approx 10^8$ cm$^{-2}$, allowing a neutral state with a few charge carriers per entire micron-scale device. Above liquid helium temperatures, the electronic properties of such devices are intrinsic, being governed by thermal excitations only. This yields that the Dirac point can be approached within 1 meV, a limit currently set by the remaining charge inhomogeneity. No sign of an insulating state is observed down to 1 K, which establishes the upper limit on a possible bandgap.

Graphene has a unique conical spectrum and its electronic properties at low energies $E$ are often described by a two-dimensional Dirac-like equation.[1,2] The resulting difference from the conventional electronic systems becomes most prominent and interesting near zero $E$ where graphene's Fermi surface shrinks into a point (Dirac point). Unfortunately, experimental devices are always subject to disorder, finite size and other factors limiting graphene's both quality and homogeneity. In particular, local variations of chemical doping and/or strain[3] disallow the Dirac point (DP) to be achieved uniformly over the entire device area so that neutral graphene is usually split into a system of electron-hole (e-h) puddles,[4,5] a state usually referred to as the neutrality point (NP). This charge inhomogeneity impedes investigation of graphene's intrinsic properties in the immediate proximity of the DP.

The standard devices made from graphene on $SiO_2$ typically exhibit density fluctuations $\delta n \sim 5\times10^{10}$ cm$^{-2}$ (ref 4) and field-effect mobilities $\mu \sim 10,000$ cm$^2$V$^{-1}$s$^{-1}$ (ref 1) which effectively smears the DP over $E \approx 20$ meV. Significant progress has recently been achieved by depositing graphene onto or encapsulating it within atomically flat boron nitride,[6,7] in which case $\delta n$ can reach nearly $10^9$ cm$^{-2}$ and $\mu$ up to $\approx 500,000$ cm$^2$V$^{-1}$s$^{-1}$.[7] An alternative approach is to use current-annealed suspended devices, which despite the 2-probe geometry usually exhibit $\mu$ up to 200,000 cm$^2$V$^{-1}$s$^{-1}$ and $\delta n < 10^{10}$ cm$^{-2}$.[8–11] More recently, even higher $\mu \approx 10^6$ cm$^2$V$^{-1}$s$^{-1}$ were reported in some suspended devices.[12,13] However, this quality is still lower than that of graphene crystals found on top of bulk graphite, in which case quantum mobility $\mu_Q$ was found $\approx 10^7$ cm$^2$V$^{-1}$s$^{-1}$ at a fixed carrier concentrations $n \approx 3\times10^9$ cm$^{-2}$.[14] Unfortunately, such crystals are in direct contact with graphite and do not allow one to either vary $n$ or measure transport properties at the DP.

In this Letter, we use two types of suspended monolayer graphene devices with $\mu >10^6$ cm$^2$V$^{-1}$s$^{-1}$ at $n <2\times10^{10}$ cm$^{-2}$ to address two common questions that are often posed in theory papers discussing graphene's intrinsic properties (see, e.g., refs. 15–21). How close is it possible to approach the Dirac point in state-of-the-art devices? And is there any many-body or spin-orbit bandgap? We show that the DP can be reached within 1 meV, a limit given by our devices' homogeneity $\delta n \sim 10^8$ cm$^{-2}$. At temperatures $T >10$ K, for all intents and purposes this is perfect graphene because both smearing and scattering at the DP is determined by thermal excitations only. Furthermore, devices' characteristics continue to evolve smoothly down to 1 K with conductivity $\sigma$ approaching linearly a finite value as $T$ decreases. This minimum conductivity $\sigma_{min}$ is close but still notably higher than $\sigma_{min} = 4e^2/\pi h$ predicted in the ballistic limit.[22,23] No sign of diverging resistivity $\rho$ yields a conservative estimate on any possible bandgap as <0.5 meV.

Figure 1a shows a micrograph of one of our suspended graphene devices, which was fabricated following the procedures described in refs. 8,10,11. In short, graphene was cleaved onto an oxidized Si wafer (300 nm SiO$_2$), and metallic contacts (Cr 3 nm/Au 100 nm) were deposited on top as shown in Figure 1a. Approximately a half of SiO$_2$ was etched away, allowing graphene to be suspended between the contacts whereas the remaining oxide served as a gate dielectric. In the second type of suspended devices, graphene was clamped between two metal pads to achieve better mechanical stability. To this end, graphene was transferred onto Au pads prefabricated on the Si wafer, and Cr/Au contacts were deposited on top of them, clamping graphene between two metal layers. This stopped graphene from moving and scrolling. Let us mention that the latter approach allowed us to fabricate a number of 4-probe devices with graphene crystals being etched into the proper Hall bar geometry.[24] Unfortunately, we found it impossible to current-anneal such Hall bars uniformly, and high $\mu$ were achieved only in the 2-probe geometry (Figure 1a). Accordingly, we discuss below only the latter devices. They were annealed in situ by using current densities ~1 mA/μm for non-clamped devices (whereas the clamped ones required currents <0.2 mA/μm). All our devices had width $W$ larger than their length $L$ (typically, $L \approx 2$μm), which we believe is important to provide homogenous annealing.

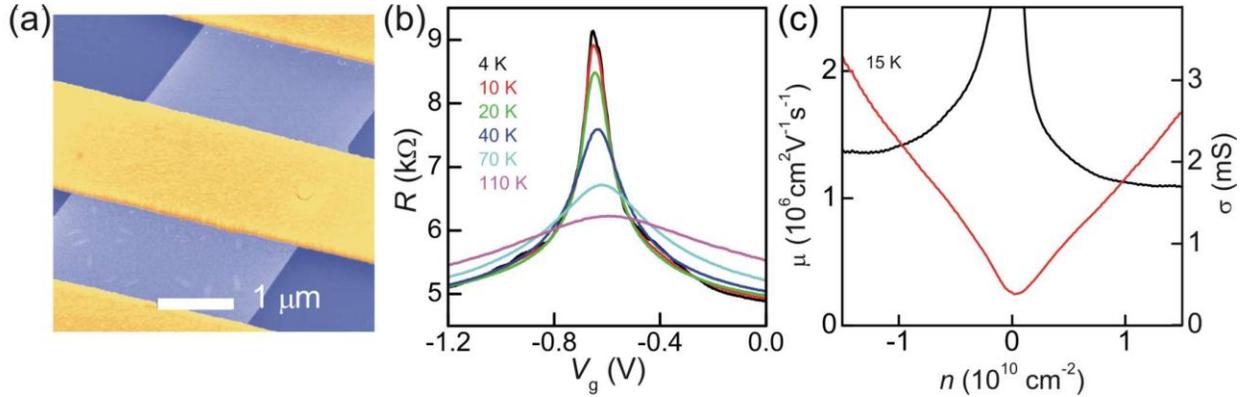

**Figure 1.** Electron transport in suspended graphene devices. (a) Scanning electron micrograph of one of our suspended devices (false colors: graphene is light blue; contacts are golden). Note some contamination over the nearest graphene region that was not annealed whereas the other, annealed region looks pristinely clean. (b) Device resistance $R$ as a function of $V_g$ at different $T$ for a clamped device. The NP occurs at $V_g = $ -0.65 V indicating electron doping. To convert $V_g$ into $n$, we have measured SdH oscillations, which yielded a constant capacitance of typically $\sim 3\times 10^{10} \times e$ cm$^{-2}$V$^{-1}$, depending on thickness of the remaining oxide ($e$ is the electron charge). (c) σ (red curve) and μ (black) for the device in (b) after subtracting $R_0 \approx 4.5$ kΩ.

Resistance $R$ as a function of applied gate voltage $V_g$ is shown in Figure 1b for one of our clamped devices. Examples for non-clamped graphene can be found in refs. 25,12,13. All our devices (>10) exhibited similar behavior and μ within a relatively narrow range of 0.5–$2\times 10^6$ cm$^2$V$^{-1}$s$^{-1}$ taken at $n = 10^{10}$ cm$^{-2}$. The only notable difference was in remnant doping Δ$n$. In non-clamped devices, $\Delta n$ was $<10^9$ cm$^{-2}$ (peak in $R(V_g)$ was centered near zero) whereas clamped devices exhibited electron doping with $\Delta n >10^{10}$ cm$^{-2}$ (Figure 1b).

To analyze the devices' quality, let us first use the standard approach assuming that electron transport in graphene is described by a combination of short- and long- range scattering mechanisms.[26,27] This leads to two terms $\rho_S$ and $\mu_L$ in graphene's resistivity $\rho = \rho_S + 1/ne\mu_L$ which are both independent on $n$ away from the regime of e-h puddles. In the 2-probe geometry, contact resistance $R_C$ must be taken into account which results in the total $n$-independent contribution $R_0 = \rho_S(L/W) + R_C$. The value of $R_C$ can accurately be estimated from the quantum Hall effect (QHE) measurements as a deviation of the 2-probe resistance from the quantized values. In our analysis, we have used $R_0$ as a single fitting parameter to obtain $\sigma(n)$ varying approximately $\propto n$ away from the NP.[26,27] An example of this linearization procedure is shown in Figure 1c. The fitting parameter $R_0$ differed from $R_C$ found from the QHE by no more than 10%. It is important to note that by definition $\mu = \mu_L/(1+\rho_S \times ne\mu_L)$ and, in ultra-high-$\mu$ devices, $\rho_S$ rather than $\mu_L$ may in principle become the characteristic defining electronic quality. However, this is not our case because $\rho_S$ typically is $\approx 50$ $\Omega$ for lower quality graphene (see, e.g., refs. 6,27), and it is reasonable to expect weaker short-range scattering in current-annealed devices. If we take the above value of $\rho_S$ as the worst case scenario, the difference between $\mu$ and $\mu_L$ in our devices would not exceed 20% for the entire reported range of $n$. This assures that $\mu = \mu_L$ is a good approximation for our experiments.

Figure 1c shows that $\mu$ in our devices is well above $10^6$ cm$^2$V$^{-1}$s$^{-1}$. The upper boundary for $n$ here is chosen to be $1.5 \times 10^{10}$ cm$^{-2}$ because, for larger $n$, the mean free path $l$ becomes comparable to $L$ (see below) and because $R$ rapidly approaches $R_C$ (Figure 1b), which makes the fitting procedure less reliable. In addition, a contribution of $\rho_S$ may lead to a gradually increasing difference between μ and $\mu_L$ at higher $n$. As for the lower boundary, we have chosen to cut off the rapidly increasing μ at $2.5 \times 10^6$ cm$^2$V$^{-1}$s$^{-1}$. One of the reasons for this somewhat arbitrarily cutoff is that low-μ graphene on SiO$_2$ is often reported to exhibit a superficially similar divergence at the NP which is an artifact. It arises from a nearly constant σ($n$) in the regime of e-h puddles, which leads to the apparent μ = σ/$ne$ diverging as $1/n$. This is certainly not our case: the increase in μ occurs at $n > 10^9$ cm$^{-2}$, that is, far away from the e-h regime at low $T$ (see below). It is also important to mention that for μ ~$10^6$ cm$^2$V$^{-1}$s$^{-1}$ and our highest $n$, $l = (\hbar/e)\mu\sqrt{\pi n}$ reaches ≈1 μm. Additional scattering at device's boundaries may reduce the apparent μ for high $n$ as reported in ref 7. Therefore, our devices were intentionally made of several μm in size, and the condition $W > L > l$ is satisfied over the whole presented range of $n$. Another possible explanation for the non-constant μ in Figure 1c (and in previously reported suspended devices[8]) could be $R_C$ varying with $n$.[24,28–30] However, $R_C$ should probably increase near the NP, reflecting poorer contact between undoped graphene and a metal.[24,28–30] If this contribution were significant, the same analysis would yield even higher μ and, probably, lead to a stronger e-h asymmetry in μ.[24] On balance, we believe that the increase in μ by a factor of >2 (see Figure 1c) is a real effect. It may originate from a decrease in scattering efficiency for a particular type of defects. An alternative explanation consistent with the behavior of μ($n$) is the renormalization of the Fermi velocity $v_F$ that can increase by a factor of 3 in suspended graphene for the same range of $n$ (see ref. 12 and our discussion below).

Although the above analysis is widely used in literature to evaluate μ, it should be considered only as a qualitative estimate. This calls for an alternative way of quantifying graphene's quality. To this end, a good measure is the magnetic field $B_q$ at which Shubnikov-de Haas (SdH) oscillations first appear.[31,25,12,13] This value is related to $\mu_Q$ through the expression $\mu_Q B_q \equiv \omega_c \tau_q = 1$, where $\omega_c$ is the cyclotron frequency and $\tau_q$ the small-angle scattering time. The formula means that carriers can complete their cyclotron orbits before being scattered. To accurately determine $\tau_q$ it is necessary to analyze SdH oscillations' amplitude as a function of $B$.[31,32] However, because the dependence is exponential, $B_q$ can be estimated within a factor of typically 30% as a value of $B$ in which the first SdH oscillation becomes noticeable at low $T$. This rule of thumb is well known for conventional electronic systems[32] and also has been noted for graphene on $SiO_2$.[27,31]

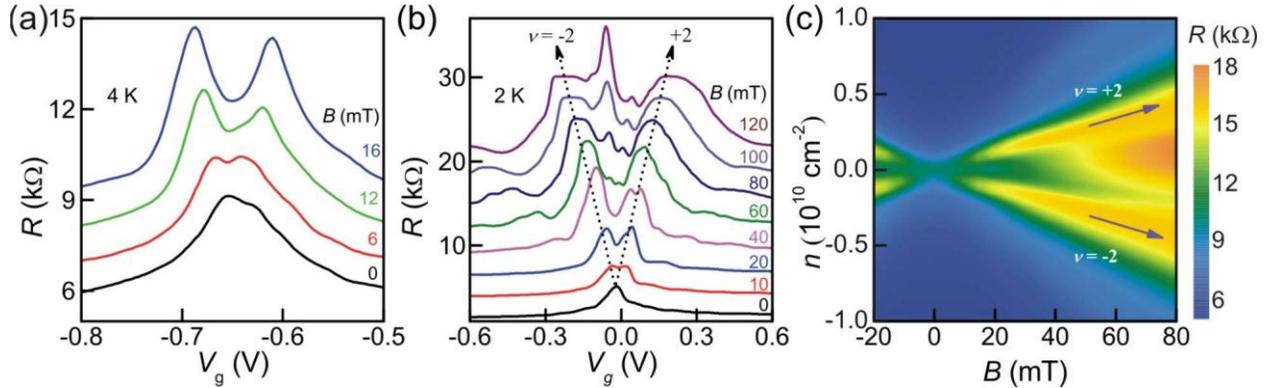

**Figure 2.** Quantum mobility. (a) $R(V_g)$ in various small $B$ for the clamped device in Figure 1. The curves are shifted vertically for clarity. (b) Similar measurements for another, non-clamped device. Splitting for $\nu = \pm 1$ is apparent already at 60 mT, and the QHE plateau $\nu = -2$ fully develops at 100 mT. (c) Color map of $R(n,B)$ for the device in (a). The arrows indicate $\nu = \pm 2$. The bright central region at zero $n$ shows that a gap is opened by $B > 50$ mT. Five times higher $B$ were necessary to open this gap for the device in (b).

As an explicit measure of $B_q$ in our experiments, we have chosen the field at which both SdH maxima appear at filling factors $\nu = \pm 2$, which results in a minimum in $R(V_g)$ at the NP (Figure 2). One can see the development of such a minimum in Figures 2a and 2b. For these two devices, the minimum appears in $B_q \approx 4$ and 6 mT, which yields $\mu_Q \approx 2.5$ and $1.6 \times 10^6$ cm$^2$V$^{-1}$s$^{-1}$, respectively (also, see Supporting Information). The former value is in good agreement with transport µ found in Figure 1c for the same device and same $n$. Note that the described procedure of defining $\mu_Q$ has previously been verified by using the 4-probe geometry for low-µ graphene[31] and, also, for encapsulated graphene.[7] In both cases, transport and quantum µ were found to agree within a factor of 2. From a theoretical point of view, $\mu_Q$ is limited by small angle scattering that destroys coherence along cyclotron orbits[33] but is insufficient for reversing the momentum's direction. Therefore, µ must be $\geq \mu_Q$ with the equality referring to the case of large-angle scatterers such as e.g. vacancies. Also, note that our typical values of $B_q$ correspond to the main cyclotron gap $\Delta(B) \approx 20–40$ K, larger than $T$ employed in the measurements as required. If we were to use higher $T$, the minimum at the NP becomes smeared at $T \approx \Delta/3$. If we were to use lower $T$, mesoscopic fluctuations (due to interference or spatial quantization) start obscuring nascent SdH oscillations (see the right shoulder in zero $B$ in Figure 2a and the 4K curve in Fig. 1b). Accordingly, for our highest-µ devices the inferred $\mu_Q$ should be considered as the lower estimate.

The onset of SdH oscillations provides a convenient way of assessing graphene's quality and, for a given $n$, yields the same electronic quality as field-effect measurements in zero $B$. However, a wider use of this analysis is difficult. In high-µ devices, it limits the determination of $\mu_Q$ to very low $n$ whereas µ can depend on $n$ (see Figure 1c). Indeed, in our suspended devices SdH oscillations appear below 10 mT, that is, at $n \sim 10^9$ cm$^{-2}$. To relate quantum and transport µ over a wider range on $n$ would require measurements of $\rho(B)$ at different $n$. This presents another problem because in ultra-high-µ devices a gap can open at the DP in $B$ as low as 50 mT (Figure 2c). The gap results in an exponential increase in $\rho$ and can overwhelm nascent SdH peaks. Despite these limitations, $\mu_Q$ refers to a real physical phenomenon (onset of Landau quantization) and, therefore, provides a sensible measure of graphene's quality especially for the 2-probe geometry, in which the standard approach of defining the field-effect μ has many other limitations discussed above.

The appearance of SdH oscillations in $B$ <10 mT and $n$ <$10^9$ cm$^{-2}$ sets up an energy scale of ≈3 meV, at which graphene's intrinsic properties are not smeared and can be probed by magnetotransport (see, e.g., measurements of $v_F$ in ref 12). To approach even closer to the DP, we analyze the behavior of σ($n$) in zero $B$. Figure 1b shows that the resistance peak continues to sharpen down to our lowest $T$ and, in some devices, we found it smeared along the $n$-axis on a scale of only ~$10^8$ cm$^{-2}$ (Figure 3a). This corresponds to $E$ ≈1 meV and implies that graphene's conductance at the NP is provided by one electron per square micron or several per the entire device. To our knowledge, every other material exhibits an insulating behavior at such low $n$. Moreover, the energy separation between spatially quantized levels is δ$E$ =2$hv_F/W$ ≈2meV and it is surprising that quantization effects remain so small: Our devices exhibit conductance fluctuations (e.g., 4K curve in Figure 1b) but no sign of an insulating state (Figure 3b). To this end, we note that decoherence is expected to increase near the DP[1,12] which may suppress quantization effects. To avoid confusion, let us mention that a metal-insulator transition was previously reported for encapsulated graphene, and this was explained by Anderson localization.[34] We attribute the difference between suspended and encapsulated graphene to a different density of intervalley scatterers, which presence is essential for localization.[1,34]

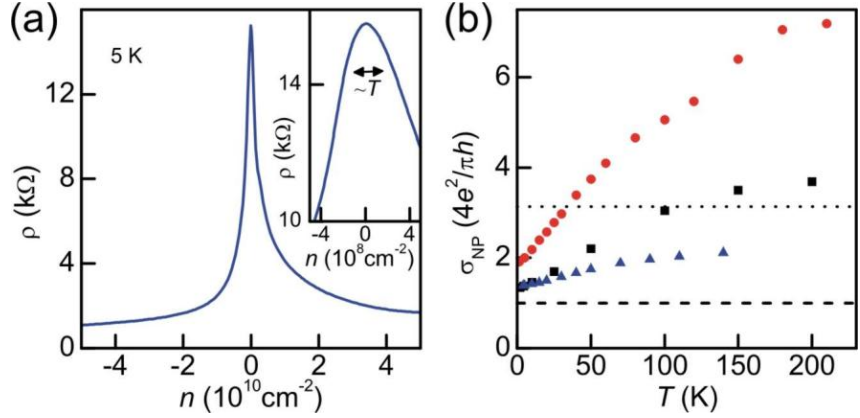

**Figure 3.** Height and broadening of the resistivity peak. (a) $\rho(n)$ for one of our devices. Inset: thermal smearing of the tip. (b) $T$ dependence of $\sigma$ at the NP for 3 devices. The dashed line indicates the ballistic limit; the dotted one is $4e^2/h$. For the sake of generality, no contact resistance is subtracted, which would result in slightly higher values of $\sigma_{NP}$.

Figure 3b shows $\sigma$ at the NP ($\sigma_{NP}$) as a function of $T$ for three monolayer devices. $\sigma_{NP}$ increases linearly with $T$ below 100 K and tends to saturate at higher $T$. The slope of the increase varies from sample to sample. This behavior is in agreement with the earlier reports on suspended graphene[35,8] and a recent theoretical model.[36] The latter suggests $\sigma_{NP} \propto T/\Gamma$ where $\Gamma$ is the broadening of the DP due to short-range scattering. At higher $T$, thermal excitations lead to additional scattering and, therefore, saturation in $\sigma_{NP}(T)$. The steepest slope in Figure 3b is for the device with the highest $\mu$ which is consistent with the model. Furthermore, $\sigma_{NP}$ in the low-$T$ limit ($\sigma_{min}$) is notably lower than $\sim 4e^2/h$, the typical value of $\sigma_{min}$ for graphene on $SiO_2$ (dotted line in Figure 3b) but still higher than $\sigma_{min} = 4e^2/\pi h$ expected in the ballistic limit (dashed).[22,23] The missing factor of $\pi$ in $\sigma_{min}$ of low-$\mu$ graphene is consensually attributed to the presence of e-h puddles.[37–39] Our suspended devices exhibit little inhomogeneity, and the observed deviations from the ballistic limit are probably due to scattering at the contact interface.[40]

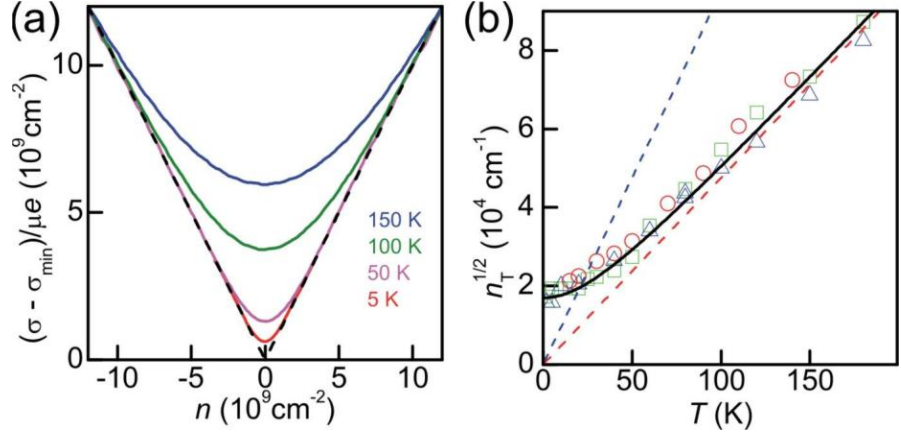

**Figure 4.** Thermal broadening of the DP. (a) The number of carriers in graphene, $N$, as a function of gate-induced $n$ at different $T$. The dashed line shows $N = n$, and the solid curves are measurements. (b) Smearing of the DP for the same 3 samples as in Figure 3b. Symbols are the experiment; the curves are fits to the theoretical expression[13] $n_T = (\pi/6)(T/v_F\hbar)^2 + \delta n$ where $\hbar$ is the reduced Planck constant. The blue and red dashed lines are for zero $\delta n$ and $v_F = 1\times$ and $2\times 10^6$ ms$^{-1}$, respectively. The solid curve takes into account a finite $\delta n$.

In our suspended graphene with little inhomogeneity, it is instructive to analyze the thermal broadening of the DP. This is a qualitative effect, easily seen on the raw curves (see Fig. 1b and Supporting Information). To introduce some quantitative measure of the observed broadening, we suggest the following scheme. In Figure 4a, we have assumed that the observed $\sigma_{min}$ describes the limit of zero density and, therefore, the number of charge carriers can be evaluated as $N = (\sigma - \sigma_{min})/\mu e$. At zero $T$ and in perfect graphene, $N$ should be equal to $n$ induced by $V_g$, as shown by the dashed lines in Figure 4a. After taking into account different $\mu$ for electrons and holes, the experimental data fall nicely on the plotted linear dependence $N = n$ if the Fermi energy $>T$. Close to the DP, $T$ induces additional carriers. The experimental curves in Figure 4a allow accurate fits by the equations derived in ref 13. To simplify the analysis, one can also employ the analytical expression $\sigma(T) \propto n_T(1+0.11n^2/n_T^2)$ valid for $n < n_T$. From these fits we determine the number of thermal-excited carriers at the DP, $n_T$, which presents a quantitative measure for the DP broadening at different $T$ (see Supplementary Information and ref. 13).

Figure 4b plots $n_T(T)$ for 3 different devices. At high $T$, the number of thermal excitations is sample independent and evolves $\propto T^2$, as expected due to the linear density of states.[26,13] We emphasize that this result is independent of the described model and will be the same for any chosen measure of the DP broadening. However, due to good agreement between our experiment and the theory[13], we can take a step further and analyze the absolute value of the broadening. If we use the standard value of $v_F = 1\times10^6$ ms$^{-1}$, the theory predicts ~4 times more carriers than observed experimentally (Figure 4b). To explain this disagreement, we recall that the Fermi velocity is renormalized at low energies. The best fit to our experimental data using a constant $v_F$ yields $\approx 2\times10^6$ ms$^{-1}$, which is in agreement with the previously observed values of $v_F$ for the same range of $n$ (ref 12) and consistent with the increase in $\mu$ in Figure 1c. The experimental data for $\mu(n)$ and $n_T(T)$ in Figures 1c and 4b also allow sensible fits by using the logarithmically diverging $v_F(n)$[12] but in our opinion this extension goes beyond the accuracy of our experiment and the reliability of theoretical assumptions (Supplementary Information).

In the limit of low $T$, the experimental dependences $n_T$ intersect the y-axis, and this yields $\delta n$. For the devices in Figure 4b, $\delta n \approx 2$–$4\times10^8$ cm$^{-2}$. Such a high homogeneity is surprising and difficult to understand because metal contacts should result in a significant charge transfer into graphene. This should make graphene strongly doped near the Au contacts[40–42,24,30,28] and, therefore, result in high charge inhomogeneity along the 2-probe device. In contrast, remnant doping $\Delta n$ is usually tiny for conventional (non-clamped) devices.[8,10,11] This can be explained by the Cr sublayer that is probably oxidized and effectively decouples graphene from the top Au but nevertheless provides sufficiently low $R_C$. Therefore, there is no contradiction as both $\Delta n$ and $\delta n$ are small for non-clamped devices. The situation is different for the case of clamped devices in which $\Delta n > 10^{10}$ cm$^{-2}$ but $\delta n$ is still $\sim 10^{8}$ cm$^{-2}$. This shows that the charge transfer does occur but is highly uniform so that $n$ varies little ($\leq 1\%$) over the devices' entire length. We speculate that the homogeneous doping is related to the fact that our devices are nearly ballistic.[40] The observation requires further theoretical analysis which is beyond the scope of the present paper.

To conclude, we have answered the two questions posed in the introduction. By using suspended high-µ devices, it is possible to approach the Dirac point within 1 meV, and there is no bandgap in graphene larger than 0.5 meV. Some features in the reported behavior such as thermal smearing and an increase in µ near the DP are consistent with the previously reported renormalization of the Fermi velocity at low energies.

AUTHOR INFORMATION

**Corresponding Author**

*E-mail: mayorov@gmail.com.

ACKNOWLEDGMENT

We thank M. Katsnelson and M. Fogler for useful discussions.

REFERENCES

(1) Castro Neto, A. H.; Guinea, F.; Peres, N. M. R.; Novoselov, K. S.; Geim, A. K. The electronic properties of graphene. *Rev. Mod. Phys.* **2009**, *81*, 109–162.

(2) Peres, N. M. R. Colloquium: The transport properties of graphene: An introduction. *Rev. Mod. Phys.* **2010**, *82*, 2673–2700.

(3) Gibertini, M.; Tomadin, A.; Guinea, F.; Katsnelson, M. I.; Polini, M. Electron-hole puddles in the absence of charged impurities. Preprint arXiv:1111.6280, 2012.

(4) Martin, J.; Akerman, N.; Ulbricht, G.; Lohmann, T.; Smet, J. H.; von Klitzing, K.; Yacoby, A. Observation of electron–hole puddles in graphene using a scanning single-electron transistor. *Nature Phys.* **2008**, *4*, 144–148.

(5) Xue, J.; Sanchez-Yamagishi, J.; Bulmash, D.; Jacquod, P.; Deshpande, A.; Watanabe, K.; Taniguchi, T.; Jarillo-Herrero P.; LeRoy, B. J. Scanning tunnelling microscopy and spectroscopy of ultra-flat graphene on hexagonal boron nitride. *Nature Mater.* **2011**, *10*, 282–285.

(6) Dean, C. R.; Young, A. F.; Meric, I.; Lee, C.; Wang, L.; Sorgenfrei, S.; Watanabe, K.; Taniguchi, T.; Kim, P.; Shepard, K. L.; Hone, J. Boron nitride substrates for high-quality graphene electronics. *Nature Nanotech.* **2010**, *5*, 722–726.

(7) Mayorov, A. S.; Gorbachev, R. V.; Morozov, S. V.; Britnell, L.; Jalil, R.; Ponomarenko, L. A.; Blake, P.; Novoselov, K. S.; Watanabe, K.; Taniguchi, T.; Geim, A. K. Micrometer-scale ballistic transport in encapsulated graphene at room temperature. *Nano Lett.* **2011**, *11*, 2396–2399.

(8) Du, X.; Skachko, I.; Barker, A.; Andrei, E. Y. Approaching ballistic transport in suspended graphene. *Nature Nanotech.* **2008**, *3*, 491–495.

(9) Bolotin, K. I.; Ghahari, F.; Shulman, M. D.; Stormer, H. L.; Kim, P. Observation of the fractional quantum Hall effect in graphene. *Nature* **2009**, *462*, 196–199.

(10) Bolotin, K. I.; Sikes, K. J.; Jiang, Z.; Klima, M.; Fudenberg, G.; Hone, J.; Kim, P.; Stormer, H. L. Ultrahigh electron mobility in suspended graphene. *Solid State Commun.* **2008**, *146*, 351–355.

(11) Bao, W.; Velasco Jr, J.; Zhang, F.; Jing, L.; Standley, B.; Smirnov, D.; Bockrath, M.; MacDonald, A.; Lau, C. N. Minimum conductivity and evidence for phase transitions in ultra-clean bilayer graphene. Preprint arXiv:1202.3212, 2012.

(12) Elias, D. C.; Gorbachev, R. V.; Mayorov, A. S.; Morozov, S. V.; Zhukov, A. A.; Blake, P.; Ponomarenko, L. A.; Grigorieva, I. V.; Novoselov, K. S.; Guinea F.; Geim, A. K. Dirac cones reshaped by interaction effects in suspended graphene. *Nature Phys.* **2011**, *7*, 701–704.

(13) Mayorov, A. S.; Elias, D. C.; Mucha-Kruczynski, M.; Gorbachev, R. V.; Tudorovskiy, T.; Zhukov, A.; Morozov, S. V.; Katsnelson, M. I.; Fal'ko, V. I.; Geim, A. K.; Novoselov, K. S. Interaction-driven spectrum reconstruction in bilayer graphene. *Science* **2011**. *333*, 860-863.

(14) Neugebauer, P.; Orlita, M.; Faugeras, C.; Barra, A.; Potemski, M. How perfect can graphene be? *Phys. Rev. Lett.* **2009**, *103*, 136403.

(15) Khveshchenko, D. V. Ghost excitonic insulator transition in layered graphite. *Phys. Rev. Lett.* **2001**, *87*, 246802.

(16) Drut, J. E.; Lähde, T. A. Is graphene in vacuum an insulator? *Phys. Rev. Lett.* **2009**, *102*, 026802.

(17) Liu, G., Li, W.; Cheng, G. Interaction and excitonic insulating transition in graphene. *Phys. Rev. B* **2009**, *79*, 205429.

(18) Gamayun, O. V.; Gorbar, E. V.; Gusynin, V. P. Gap generation and semimetal-insulator phase transition in graphene. *Phys. Rev. B* **2010**, *81*, 075429.

(19) Yao, Y.; Ye, F.; Qi, X.; Zhang, S.; Fang, Z. Spin-orbit gap of graphene: First-principles calculations. *Phys. Rev. B* **2007**, *75*, 041401.

(20) Kane, C. L.; Mele, E. J. Quantum spin Hall effect in graphene. *Phys. Rev. Lett.* **2005**, *95*, 226801.

(21) Herbut, I. F.; Juričić, V.; Vafek, O. Relativistic Mott criticality in graphene. *Phys. Rev. B* **2009**, *80*, 075432.

(22) Katsnelson, M. I. Zitterbewegung, chirality, and minimal conductivity in graphene. *Eur. Phys. J. B* **2006**, *51*, 157–160.

(23) Tworzydło, J.; Trauzettel, B.; Titov, M.; Rycerz, A.; Beenakker, C. W. J. Sub-Poissonian shot noise in graphene. *Phys. Rev. Lett.* **2006**, *96*, 246802.

(24) Blake, P.; Yang, R.; Morozov, S. V.; Schedin, F.; Ponomarenko, L. A.; Zhukov, A. A.; Nair, R. R.; Grigorieva, I. V.; Novoselov, K. S.; Geim, A. K. Influence of metal contacts and charge inhomogeneity on transport properties of graphene near the neutrality point. *Solid State Commun.* **2009**, *149*, 1068–1071.

(25) Castro, E. V.; Ochoa, H.; Katsnelson, M. I.; Gorbachev, R. V.; Elias, D. C.; Novoselov, K. S.; Geim, A. K.; Guinea, F. Limits on charge carrier mobility in suspended graphene due to flexural phonons. *Phys. Rev. Lett.* **2010**, *105*, 266601.

(26) Hwang, E. H.; Adam, S.; Das Sarma, S. Carrier transport in two-dimensional graphene layers. *Phys.Rev. Lett.* **2007**, *98*, 186806.

(27) Morozov, S. V.; Novoselov, K. S.; Katsnelson, M. I.; Schedin, F.; Elias, D. C.; Jaszczak, J. A.; Geim, A. K. Giant intrinsic carrier mobilities in graphene and its bilayer. *Phys. Rev. Lett.* **2008**, *100*, 016602.

(28) Lee, E. J. H.; Balasubramanian, K.; Weitz, R. T.; Burghard, M.; Kern, K. Contact and edge effects in graphene devices. *Nature Nanotech.* **2008**, *3*, 486–490.

(29) Wu, Y.; Perebeinos, V.; Lin, Y.; Low, T.; Xia, F.; Avouris, P. Quantum Behavior of Graphene Transistors near the Scaling Limit. *Nano Lett.* **2012**, *12*, 1417–1423.

(30) Hannes, W.; Jonson, M.; Titov, M. Electron-hole asymmetry in two-terminal graphene devices. *Phys. Rev. B* **2011**, *84*, 045414.

(31) Monteverde, M.; Ojeda-Aristizabal, C.; Weil, R.; Bennaceur, K.; Ferrier, M.; Guéron, S.; Glattli, C.; Bouchiat, H.; Fuchs, J. N.; Maslov, D. L. Transport and elastic scattering times as probes of the nature of impurity scattering in single-layer and bilayer graphene. *Phys. Rev. Lett.* **2010**, *104*, 126801.

(32) Lonzarich, G. G. *Electrons at the Fermi surface, edited by M. Springford*, (Cambridge University, Cambridge, 1980), Chap. 6.

(33) Harrang, J. P.; Higgins, R. J.; Goodall, R. K. Quantum and classical mobility determination of the dominant scattering mechanism in the two-dimensional electron gas of an AlGaAs/GaAs heterojunction. *Phys. Rev. B* **1985**, *32*, 8126–8135.

(34) Ponomarenko, L. A.; Geim, A. K.; Zhukov, A. A.; Jalil, R.; Morozov, S. V.; Novoselov, K. S.; Grigorieva, I. V.; Hill, E. H.; Cheianov, V. V.; Fal'ko, V. I.; Watanabe, K.; Taniguchi T.;

Gorbachev, R. V. Tunable metal–insulator transition in double-layer graphene heterostructures. *Nature Phys.* **2011**, *7*, 958–961.

(35) Bolotin, K. I.; Sikes, K. J.; Hone, J.; Stormer, H. L.; Kim, P. Temperature-dependent transport in suspended graphene. *Phys. Rev. Lett.* **2008**, *101*, 096802.

(36) Sentef, M.; Kollar, M.; Kampf, A. P. DC conductivity of graphene with disorder. Preprint arXiv:1203.2216, 2012.

(37) Cheianov, V. V.; Fal'ko, V. I.; Altshuler, B. L.; Aleiner, I. L. Random resistor network model of minimal conductivity in graphene. *Phys. Rev. Lett.* **2007**, *99*, 176801.

(38) Adam, S.; Hwang, E. H.; Galitski, V. M.; Das Sarma S. A self-consistent theory for graphene transport. *Proc. Natl. Acad. Sci.* **2007**, *104*, 18392–18397.

(39) Fogler, M. M. Neutrality point of graphene with coplanar charged impurities. *Phys. Rev. Lett.* **2009**, *103*, 236801.

(40) Golizadeh-Mojarad, R.; Datta, S. Effect of contact induced states on minimum conductivity in graphene. *Phys. Rev. B* **2009**, *79*, 085410.

(41) Xia, F.; Perebeinos, V.; Lin, Y.; Wu, Y.; Avouris. P. The origins and limits of metal–graphene junction resistance. *Nature Nanotech.* **2011**, *6*, 179–184.

(42) Khomyakov, P. A.; Giovannetti, G.; Rusu, P. C.; Brocks, G.; van den Brink, J.; Kelly, P. J. First-principles study of the interaction and charge transfer between graphene and metals. *Phys. Rev. B* **2009**, *79*, 195425.

## ASSOCIATED CONTENT

**Transport measurements in magnetic field**

Our transport measurements were carried out in two He$^4$-cryostats (equipped with superconducting magnets of 12 and 16 T) at temperatures from 1.5 K to 210 K by using the standard low-frequency (30.5 Hz) lock-in technique. In order to avoid an overheating of the electron system, excitation currents between 10 and 100 nA were applied. The residual magnetic fields were typically ~5 mT and ~14 mT for 12 and 16 T magnets, respectively. This offset was taken into account so that the behaviour was symmetric with respect to ±$B$ (Figure S1). We note that, despite being relatively small and often neglected, such remnant $B$ correspond to a cyclotron gap of a few meV, and ignoring this effect in ultra-high-μ graphene may lead to artefacts.

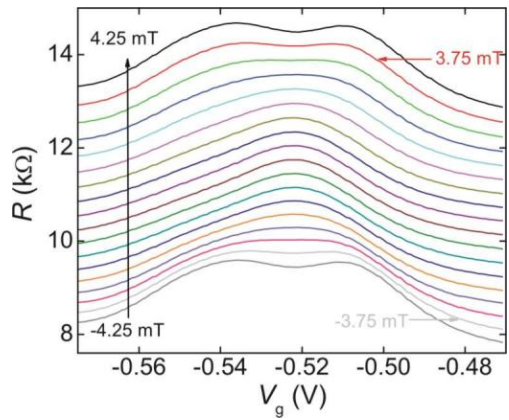

Figure S1. Resistance as a function of gate voltage for several $B$ varying from -4.25 mT to 4.25 mT. The curves are shifted vertically for clarity, and the onset of SdH oscillations is indicated by the arrows.

Figure S1 provides yet another example of the emergence of SdH oscillations in a clamped device. The measurements were performed at 4 K by using a 10 nA current. The splitting near the DP becomes discernible at ±3.75 mT. This is the smallest $B_q$ we have so far observed in both clamped and unclamped devices. It corresponds to $\mu_Q \approx 2.8 \times 10^6$ cm$^2$V$^{-1}$s$^{-1}$.

**Dirac point broadening analysis**

We have used the standard transport expression for graphene, which assumes short and long range scattering: $\rho=\rho_0+1/en\mu$. This expression follows from the semiclassical Boltzmann equation, which validity limits are discussed in detail in refs. S1,S2. In order to use the Boltzmann equation near the DP, it requires the condition $2\pi\varepsilon_F\tau/h \gg 1$, where $\varepsilon_F$ is the Fermi energy, $\tau$ the transport scattering time, and $h$ the Plank constant. This condition can be re-written as $2\pi l \gg \lambda_F$, where $l$ is the mean free path and $\lambda_F$ the Fermi wavelength. The expression is valid in our experiments, except for very low concentrations where the Fermi wavelength reaches the sample size.

In order to find the concentration of the thermally-excited carriers $n_T$ we have used the theory presented in ref 13. Its main assumption is that conductivity $\sigma$ of graphene at zero $T$ takes the form $\sigma_0 = \beta n$. This expression is known to be a good approximation in the Fermi liquid regime $\varepsilon_F > T$ (with $\beta = e\mu$) but its validity remains untested near the DP where the broadening takes place. For example, the theory[13] does not take into account that $\mu$ can depend on $T$ and $v_F$ can change due to many-body effects. However, we expect this remains a reasonable approximation if $\beta$ remains nearly constant in the studied density range.

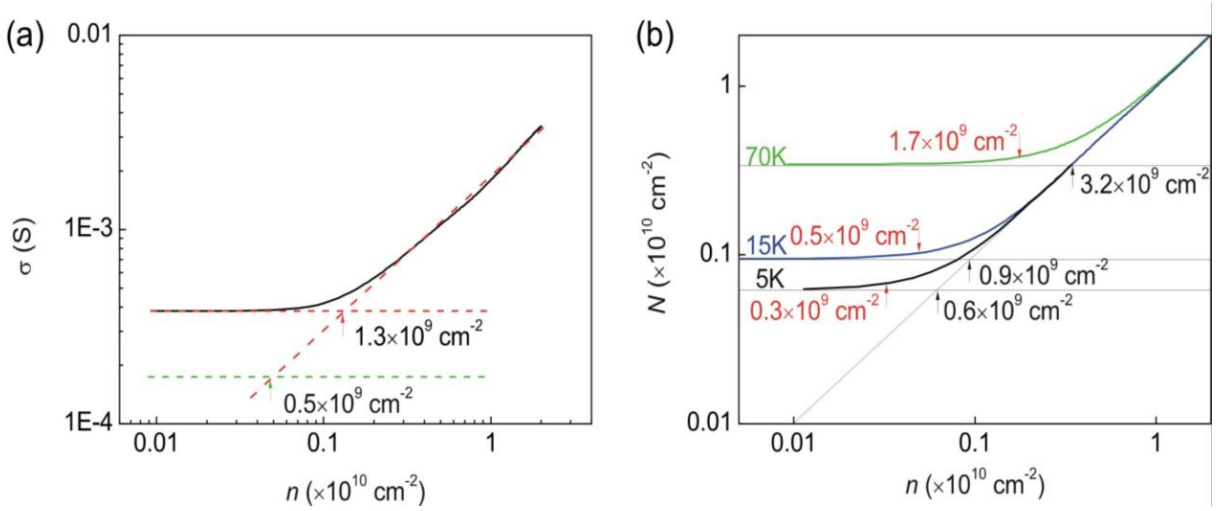

Figure S2. (a) Conductivity as a function of concentration at 15K. The red arrow shows one of possible ways to define the thermal broadening of the DP. The green arrow indicates $n_T$ expected in theory. (b) Number of carriers in graphene as a function of $n$ for $T =5$, 15 and 70K. The black arrows mark the DP broadening using the same definition as in (a). The red arrows show $n_T$ found from our theoretical fits.

The DP broadening depends on both disorder and temperature. At low enough $T$ the broadening is determined by disorder only. Figure S2a plots the curve of Figure 1c in a logarithmic scale. One can see that $\sigma$ stops changing with $n$ at $\approx 1 \times 10^9 \text{cm}^{-2}$. The DP broadening can be estimated as the intersection of the asymptotic lines at high and low concentrations as shown in Fig. S2a. This approach can apparently be used to quantify the broadening on the experimental curves. However, in order to take into account changes in $\mu$ for different devices, we have chosen to analyze $N = \sigma/e\mu$ (see Fig. 2Sb). At high $T$, we have found the behavior of $N$ sample-independent. If the DP broadening is again defined as the intersection of the corresponding lines, Figure S2b shows that this representation leads to slightly different absolute values (cf. Figs. S2a,b).

To achieve a better-than-logarithmic accuracy in defining the DP broadening, we employ the theory developed in ref 13, which involves the full fitting of curves $N(n)$ as discussed below. At finite $T$, the Fermi energy $\xi$ is a function of $n$ and $T$:

$$n(\xi) = \frac{12}{\pi^2} n_T \sinh\left(\frac{\xi}{kT}\right) \int_0^\infty \frac{x\,dx}{\cosh x + \cosh\left(\frac{\xi}{kT}\right)},$$

where $n_T = \frac{\pi T^2}{6(\hbar v_F)}$ is the number of thermally excited electrons (holes). $n_T$ is the only adjustable parameter that depends on $v_F$. The number of charge carriers is then determined by[13]

$$N(n) = \frac{12}{\pi^2} n_T \int_0^\infty \frac{x e^{-x} dx}{chx + ch\left(\frac{\xi}{kT}\right)} + n \coth\left(\frac{\xi}{kT}\right),$$

which was used to fit the experimental curves in Figure 4a. The example of the fit is shown in Figure S3 for the 70K curve of Figure S2b.

Instead of using the two equations presented above (the fit involves numerical evaluation of the integrals) we can use a simpler equation. In the case of $n<n_T$, the concentration of the charge carriers as a function of $T$ and $n$ is given by[13]

$$N = \sigma(T,n)/\beta \approx 2n_T(T)\left[1 + \frac{\pi^2}{192\ln^2(2)} \frac{n^2}{n_T^2}\right].$$

Note that the used equations depend only on $T$ and $n$ and are "thermodynamic" values, that is, do not contain scattering rates which determine temperature dependence of $\sigma$.

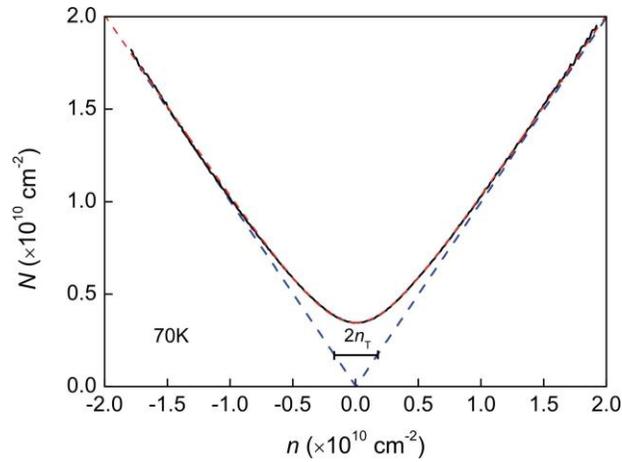

Figure S3. Concentration of the charge carriers in graphene, *N*, as a function of the gate-induced concentration *n*. The solid black line is obtained from the resistance measurements at 70 K. The red dotted line is the best fit with $n_T = 1.7 \times 10^9$ cm$^{-2}$. The blue dashed line represents an ideal graphene at zero temperature.

(S1) Katsnelson, M. I. *Graphene: Carbon in Two Dimensions,* (Cambridge University, Cambridge, 2012), Chap. 11.

(S2) Auslender. M.; Katsnelson, M. I. Generalized kinetic equations for charge carriers in graphene *Phys. Rev. B* **2007**, *76*, 235425.